\documentclass[useAMS,usenatbib]{mn2e}
\usepackage{epsfig}
\usepackage{amsmath}

\newcommand{\be}{\begin{equation}}
\newcommand{\beq}{\begin{equation}}
\newcommand{\ba}{\begin{eqnarray}}
\newcommand{\ee}{\end{equation}}
\newcommand{\eeq}{\end{equation}}
\newcommand{\ea}{\end{eqnarray}}

\newcommand{\hs}{\hspace{1mm}}

\newcommand{\apj}{ApJ}
\newcommand{\aap}{A\&A}
\newcommand{\apjl}{ApJL}
\newcommand{\mnras}{MNRAS}
\newcommand{\aj}{AJ}
\newcommand{\apjs}{ApJS}
\newcommand{\nat}{{\it Nature}}

\newcommand{\pasj}{PASJ}
\newcommand{\pasp}{PASP}
\newcommand{\procspie}{Proceedings of SPIE}

\newcommand{\llya}{$L_{{\rm Ly}\alpha}$}

\def\lsim{~\rlap{$<$}{\lower 1.0ex\hbox{$\sim$}}}

\def\gsim{~\rlap{$>$}{\lower 1.0ex\hbox{$\sim$}}}

\title[The Polarization of Scattered Ly$\alpha$ Around High-z
Galaxies]{The Polarization of Scattered Ly$\alpha$ Radiation Around
High-Redshift Galaxies}

\author[Mark Dijkstra \& Abraham Loeb]{Mark
Dijkstra\thanks{E-mail:mdijkstr@cfa.harvard.edu} and Abraham
Loeb\thanks{E-mail:aloeb@cfa.harvard.edu}\\ Harvard-Smithsonian Center for
Astrophysics, 60 Garden Street, Cambridge, MA 02138, USA}

\def\LaTeX{L\kern-.36em\raise.3ex\hbox{a}\kern-.15em
    T\kern-.1667em\lower.7ex\hbox{E}\kern-.125emX}

\begin{document}

\date{\today}
\pagerange{\pageref{firstpage}--\pageref{lastpage}} \pubyear{2007}

\maketitle

\label{firstpage}
\begin{abstract}
The high-redshift Universe contains luminous Ly$\alpha$ emitting sources
such as galaxies and quasars. The emitted Ly$\alpha$ radiation is often
scattered by surrounding neutral hydrogen atoms. We show that the scattered
Ly$\alpha$ radiation obtains a high level of polarization for a wide range
of likely environments of high-redshift galaxies. For example, the
back-scattered Ly$\alpha$ flux observed from galaxies surrounded by a
superwind-driven outflow may reach a fractional polarization as high as
$\sim 40\%$. Equal levels of polarization may be observed from neutral
 collapsing protogalaxies. Resonant scattering in the diffuse
intergalactic medium typically results in a lower polarization amplitude
($\lsim 7\%$), which depends on the flux of the ionizing background.
Spectral polarimetry can differentiate between Ly$\alpha$ scattering off
infalling gas and outflowing gas; for an outflow the polarization should
increase towards longer wavelengths while for infall the opposite is
true. Our numerical results suggest that Ly$\alpha$ polarimetry is feasible
with existing instruments, and may provide a new diagnostic of
the distribution and kinematics of neutral hydrogen around high-redshift
galaxies. Moreover, polarimetry may help suppress infrared lines originating in the Earth's atmosphere, and thus improve the sensitivity of ground-based observations to high-redshift Ly$\alpha$ emitting galaxies outside the currently available redshift windows.
\end{abstract}

\begin{keywords} 
cosmology: theory--galaxies: high redshift--line: formation--polarization--scattering
\end{keywords}
 
\section{Introduction}
\label{sec:intro}
The Lyman-$\alpha$ (Ly$\alpha$) line serves as an excellent tracer of high-redshift galaxies and quasars \citep[e.g.][]{Partridge,Hu96,Steidel96,Trager97,Rhoads00,Hu02,Ko03,Ka06,Tapken06,Westra06,Stanway07,Ouchi07,Rauch07}. Searches for redshifted Ly$\alpha$ emission have discovered galaxies robustly out to redshifts
$z=6.96$ \citep{Iye06}, and potentially out to $z=10$
\citep{Stark07}. These observations provide a unique glimpse into galaxy formation at high-redshift \citep[][]{MR02,DW07}, and into the Epoch of Reionization
\citep[EoR, e.g.][]{Loeb99,Haiman99,MR04,HC05,FL06,LF,Stark06,McQuinn07,Ota07,Ko07,F07,Mesinger07}. Future surveys intend to exploit this window further by pushing
to even higher redshifts and fainter flux levels \citep[e.g.][and references therein]{Stark06,N07}.

The interpretation of existing and future observations is complicated by
the fact that Ly$\alpha$ photons are typically scattered both in the
interstellar medium (ISM, e.g. Hansen \& Oh 2006) as well as in the
intergalactic medium (IGM, Loeb \& Rybicki 1999; Santos 2004; Dijkstra et
al 2007b). The impact of scattering on the observed Ly$\alpha$ flux and
spectrum may be derived by careful modeling of the observed Ly$\alpha$ line
profile, combined with constraints on the galaxy's stellar population
derived from its broad band colors \citep[][]{Verhamme06,Verhamme07}.
Also, observations of the Ly$\alpha$ line in local starburst galaxies can
shed light on the processes that regulate the transfer and escape of
Ly$\alpha$ from star forming galaxies \citep[][]{Kunth98,Hayes07}. Of
course, it is not obvious whether local starburst galaxies are representative
of high-redshift Ly$\alpha$ emitters (LAEs), and detailed modeling of the
line profile may be hampered by the quality of the data of high-redshift
LAEs.

In this paper, we describe a complementary approach to studying the
environments of high-redshift LAEs, which involves the polarization
properties of the scattered Ly$\alpha$ radiation. In the context of the
solar system, the polarization of resonantly scattered Ly$\alpha$ photons
from the sun by neutral hydrogen in the interplanetary medium and in the
Earth's atmosphere has been studied extensively \citep[e.g.][]{Brandt59},
occasionally by using a Monte-Carlo method to compute the Ly$\alpha$
radiative transfer \citep[][]{Modali72,Keller81}. In the cosmological
context, \citet{Lee98} showed that Ly$\alpha$ emerging from an unresolved
star-bursting galaxy may be polarized to a considerable level ($\sim 5\%)$,
and \citet{RL99} showed that scattering of Ly$\alpha$ photons from a point
source embedded in the Hubble flow of a neutral IGM would produce a
polarized Ly$\alpha$ halo around the source. Despite the emergence of
several Monte-Carlo Ly$\alpha$ radiative transfer codes in recent years
\citep[e.g.][]{Zheng02,Cantalupo05,Iro06,D06,Verhamme06,Laursen07,
Semelin07}, the polarization properties of scattered Ly$\alpha$ photons
around galaxies have not been investigated in a broader context beyond this
early work. 

The goal of this paper is to demonstrate that Ly$\alpha$ radiation may be
highly polarized around high-redshift galaxies in a broad range of cosmological circumstances. We will show that Ly$\alpha$ polarimetry may place constraints on the kinematics of
the gas through which the photons propagate. This is important: understanding the impact of scattering in both the ISM and IGM on the observed Ly$\alpha$ properties is required to fully exploit Ly$\alpha$ observations as a probe of the high-redshift Universe. Furthermore, we discuss the possibility of separating Ly$\alpha$ sources from low-redshift line emitters (such as [OII], [OIII], H$\alpha$, H$\beta$ emitters, and OH-skylines) based on polarimetry. 

The outline of this paper is as follows:
In \S~\ref{sec:lya} we describe the basic principles of Ly$\alpha$
radiative transfer and polarization, and how polarization is incorporated
in our radiative transfer calculations.
In \S~\ref{sec:result} we present our main numerical results.
In \S~\ref{sec:discuss} we discuss our results and their implications,
before presenting our final conclusions in \S~\ref{sec:conc}.
 The parameters for the background cosmology used throughout this paper are $(\Omega_m,\Omega_{\Lambda},h)=(0.24,0.76,0.73)$ \citep{wmap}.

\section{Ly$\alpha$ Scattering \& Polarization}
\label{sec:lya}
\subsection{Ly$\alpha$ Radiative Transfer Basics}
\label{sec:RT1}
We start by summarizing the basic principles of Ly$\alpha$ scattering.  We
express photon frequency $\nu$ in terms of a dimensionless variable
$x\equiv (\nu-\nu_0)/\Delta \nu_D$, where $\Delta \nu_D=v_{th}\nu_0/c$, and
$v_{th}$ is the thermal velocity of the hydrogen atoms in the gas, given by
$v_{th}=\sqrt{2k_B T/m_p}$, where $k_B$ is the Boltzmann constant, $T=10^4$
K the gas temperature, $m_p$ the proton mass and $\nu_0=2.47 \times
10^{15}$ Hz is the Ly$\alpha$ resonance frequency. For reference, the
optical depth through a column of hydrogen, $N_{\rm HI}$, for a photon in
the line center, $\tau_0$, is given by
\begin{equation}
\tau_0=5.9 \times 10^6 \Big{(}\frac{N_{\rm HI}}{10^{20}\hs{\rm cm}^{-2}} \Big{)} \Big{(} \frac{T}{10^4\hs{\rm K}}\Big{)}^{-0.5}.
\label{eq:tau0}
\end{equation}  
The optical depth for a photon at a frequency $x$ reduces to
\begin{equation}
\frac{\tau_x}{\tau_0}=\frac{a}{\pi}\int_{-\infty}^{\infty}
\frac{e^{-y^2}dy}{(y-x)^2+a^2}=
\left\{ \begin{array}{ll}
         \ \sim  e^{-x^2}& \mbox{core};\\
         \ \sim \frac{a}{\sqrt{\pi}x^2}& \mbox{wing},\end{array} 
\right. 
\label{eq:phi}
\end{equation} 
\citep[e.g.][]{RL79} where $a$ is the Voigt parameter given by $a=A_{21}/4
\pi \Delta \nu_D =4.7 \times 10^{-4} (v_{th}/13 \hs {\rm km \hs
s}^{-1})^{-1}$, where $A_{21}=6.25\times 10^8$ s$^{-1}$ is the Einstein
A-coefficient for the transition. The transition between `wing' and `core' occurs at $x\equiv x_{\rm t} \approx 3.3$ for $T=10^4$ K.

Under most astrophysical conditions, absorption of a Ly$\alpha$ photon is
followed be re--emission of a photon of the same energy in the frame of the
atom. However, in the observer's frame the Ly$\alpha$ photon's energy
before and after scattering is modified by Doppler shift due to the thermal
motion of the atom, and scattering is only 'partially' coherent in the
observer's frame. These Doppler shifts are important as they cause the
photon's frequency to change by an {\it rms} shift of $\Delta \nu_D$ per
scattering event \citep{Osterbrock62}. Therefore, Ly$\alpha$ scattering
through an optically thick medium can be described as a sequence of random
walks in both frequency and real space \citep[e.g][]{Harrington73,Neufeld90,Loeb99}. Frequency diffusion is very efficient and spatial diffusion occurs predominantly when photons are in the wing of the line profile \citep[e.g.][]{Adams72,Harrington73}. Hence, the last scattering event
occurs in the wing of the line profile for the majority of Ly$\alpha$
photons emerging from an optically thick medium
\footnote{Note that Ly$\alpha$ photons typically scatter only $\sim\tau_0$
times before emerging from a medium of optical depth $\tau_0$, as opposed
to the $\tau_0^2$-scaling, that is expected in the absence of frequency
diffusion. For scattering events that occur in the wing, a so called 'restoring force'  pushes the photon back towards the core by an average amount of $-1/x$ \citep{Osterbrock62,Adams72}.}.

Other quantities of relevance to our discussion are: ({\it i}) the
scattering phase function (also known as the anisotropy function) which
gives the probability $p(\theta)$ that the scattered photon is re--emitted
at an angle $\theta$ relative to the incoming photon; and ({\it ii}) The
degree of polarization $\Pi(\theta)$ caused by scattering which is defined
as $\Pi(\theta)\equiv \frac{I_{||}-I_{\perp}}{I_{||}+I_{\perp}}$, where
$I_{||}$ and $I_{\perp}$ are the intensities parallel and perpendicular to
the plane of scattering (defined by the wave vectors of the ingoing and
outgoing photons), as a function scattering angle $\theta$. Both
$p(\theta)$ and $\Pi(\theta)$ are different for core and wing
scattering, as we discuss in more detail below.

\subsection{Phase Function \& 
Degree of Polarization for Ly$\alpha$ Scattering}
\label{sec:RT2}

For resonant scattering the phase function and the degree of polarization
are strongly dependent on the atomic levels involved in the scattering
event, and must be calculated quantum-mechanically. For example, the
sequence of scattering events $1S_{1/2}\rightarrow 2P_{1/2}\rightarrow
1S_{1/2}$ result in an unpolarized isotropically re-emitted Ly$\alpha$ photon, while the scattering events $1S_{1/2}\rightarrow 2P_{3/2}\rightarrow
1S_{1/2}$ produce Ly$\alpha$ with a maximum polarization of $\frac{3}{7}$
\citep[e.g.][]{Hamilton47,Chandra60,Ahn02}. Here we used a notation
nL$_{\rm J}$, in which n, L and J are the principal quantum number, orbital
angular momentum number, and total angular momentum number of the hydrogen
atom involved in the scattering event, respectively. By summing over all
possible Ly$\alpha$ transitions, the phase function and degree of
polarization for resonant scattering were found to be
\citep[e.g.][]{Brandt59,Brasken98}
\begin{equation}
p(\theta)=\frac{11}{12}+\frac{3}{12}\cos^2\theta, \hspace{5mm} \Pi(\theta)=\frac{\sin^2\theta}{\frac{11}{3}+\cos^2 \theta}.
\label{eq:phasecore}
\end{equation} 
As is shown in Appendix~\ref{app:super}, this corresponds to a case of a
superposition of Rayleigh scattering plus isotropic scattering with
corresponding weights of $1/3$ and $2/3$. The phase function $p(\theta)$
satisfies $\int p(\theta)d\Omega=4\pi$.

In quantum mechanics a Ly$\alpha$ scattering event cannot go separately
through either $1S_{1/2}\rightarrow 2P_{1/2}\rightarrow 1S_{1/2}$ or
$1S_{1/2}\rightarrow 2P_{3/2}\rightarrow 1S_{1/2}$. Instead, each
scattering event is a superposition of scattering through both levels
simultaneously. \citet{Stenflo80} has shown that this introduces quantum
interference terms which cause Ly$\alpha$ wing-scattering to be described
by Rayleigh scattering for which
\begin{equation}
p(\theta)=\frac{3}{4}+\frac{3}{4}\cos^2\theta, \hspace{5mm} \Pi(\theta)=\frac{\sin^2\theta}{1+\cos^2 \theta}.
\label{eq:phasewing}
\end{equation} 
This remarkable result implies that Ly$\alpha$ wing scattering can produce
three times more polarization than Ly$\alpha$ resonant scattering
\citep{Stenflo80}.

\subsection{Incorporating Ly$\alpha$ Polarization 
in Radiative Transfer Codes}
\label{sec:RT3}
\begin{figure}
\vbox{\centerline{\epsfig{file=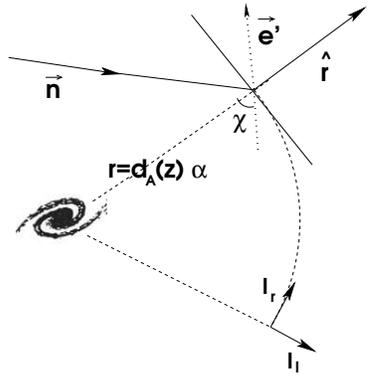,angle=0,width=4.8truecm}}}
\caption[]{Schematic geometry of the last-scattering event of a Ly$\alpha$
photon that occurs at a radius vector ${\bf r}$ away from the source
galaxy. Before last scattering the photon's propagation direction is ${\bf
n}$ (which does not necessarily lie in the plane of the sky). After
scattering, the photon travels in a direction ${\bf n}'$ which is
perpendicular to the sky plane. The polarization vector ${\bf e}'$ after
scattering lies in the plane of the sky and intersects the projected (on
the sky plane) radius vector ${\bf r}$ at an angle $\chi$. This photon
contributes $\cos^2 \chi$ and $\sin^2 \chi$ to the intensity of the
radiation field parallel ($I_l(\alpha)$) and perpendicular ($I_r(\alpha)$)
to ${\bf r}$, respectively. Here $\alpha=r/d_A(z)$ (see text for additional
details).}
\label{fig:scheme}
\end{figure}

The code we used for the Monte-Carlo Ly$\alpha$ radiative transfer is
described in \citet{D06}. The code follows individual Ly$\alpha$ photons
through spherical concentric shells with user-specified density and
velocity fields. The code accurately describes the process of frequency and
spatial diffusion in Ly$\alpha$ radiative transfer as described in
\S~\ref{sec:RT1} (including other effects that were not discussed such
as for example energy loss due to recoil). For a detailed description of the code, the interested reader is referred to \citet{D06}. The code was adapted to
the context of this paper as follows:

\begin{itemize}

\item The modified code uses different phase functions for scattering in
the core and wing as described in \S~\ref{sec:RT2}. The original code
(which was constructed for a different purpose) assumed only the Rayleigh
scattering phase function.

\begin{figure}
\vbox{\centerline{\epsfig{file=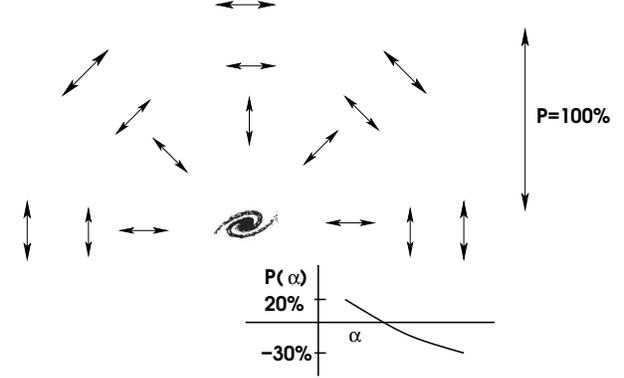,angle=0,width=8.0truecm}}}
\caption[]{Schematic illustration of the polarization of scattered
Ly$\alpha$ radiation. Arrows pointing away (towards) the galaxy represent a
radiation field for which $I_{\rm l} < I_{\rm r}$
(i.e. $\mathcal{P}(\alpha)>0$), while $I_{\rm l} > I_{\rm r}$ is
represented by arrows tangential to spheres of constant radius. The
magnitude of the polarization is represented by the size of the arrows. The
arrows vanish for unpolarized radiation. In all cases considered in this
paper, $\mathcal{P}(\alpha)<0$ (this corresponds to a Stokes parameter
$Q<0$), and the polarization vectors form concentric shells surrounding the central source.}
\label{fig:scheme2}
\end{figure}

\item In \S~\ref{sec:RT2} we showed that the polarization of scattered
Ly$\alpha$ photons depends strongly on whether Ly$\alpha$ was resonantly
scattered or not. We also mentioned in \S~\ref{sec:RT1} that the transition
between resonant and wing scattering occurs at $x\sim 3.3$. However, it is
possible to determine whether a photon scattered resonantly for each
scattering event directly from the Monte-Carlo simulation. While generating
the velocity of the atom that scatters the Ly$\alpha$ photon, a scattering
event is defined to be resonant if it occurred less than $x_{\rm crit}=0.2$
Doppler widths away from resonance {\it in the frame of the atom involved
in the scattering event} (see Appendix~\ref{app:super} for a justification
of this number). The precise choice of $x_{\rm crit}$ does not affect our
main results.

\item The code was modified to include polarization based on the scheme
developed by \citet{RL99}. Although their method is applicable strictly to
wing/Rayleigh scattering, only minor modifications are required to
calculate polarization for resonantly scattered Ly$\alpha$ photons. This is
mainly because resonant scattering is a superposition of isotropic and
Rayleigh scattering. We will briefly discuss the \citet{RL99} method, before
describing our modifications.

\citet{RL99} assigned 100\% linear polarization to individual photons by
endowing each Ly$\alpha$ photon with a polarization vector ${\bf e}$, which
is perpendicular to the photon's propagation direction
 ${\bf n}$, i.e. ${\bf e \cdot n}=0$. In this formulation, the Stokes parameters result from binning together multiple independent photons. 

The propagation direction (${\bf n'}$) of a photon after scattering is obtained by means of a rejection method: A random direction ${\bf n'}$, and a random number $R\in [0,1]$ are generated. The new propagation direction is accepted when $R<1-({\bf e \cdot n'})^2$ (as shown in Appendix~\ref{app:phase}, this corresponds to a more general case of the phase-function $p(\theta)$ given in Eq~\ref{eq:phasewing}). The polarization vector after scattering (${\bf e'}$) is obtained by normalizing the vector ${\bf g'}$, which is obtained by projecting the polarization vector prior to scattering (${\bf e}$) onto the plane normal to ${\bf n'}$. Symbolically, ${\bf e}'={\bf g}/|{\bf g}|$, where ${\bf g}={\bf e}-({\bf e \cdot n' }){\bf n'}$. 

The observed radiation is characterized by its intensities parallel ($I_l$) and perpendicular ($I_{r}$) to the radius vector to the location of last scattering, denoted by ${\bf r}$, projected on the plane of the sky (see Figure~\ref{fig:scheme}), denoted by ${\bf \hat{r}}$. For the last scattering event ${\bf n'}$ is perpendicular to the plane of the sky, and the polarization vector ${\bf e}'$ lies in the plane of the sky (see Fig~\ref{fig:scheme}). We use $\chi$ to denote the angle between the polarization vector ${\bf e}'$ and ${\bf \hat{r}}$. A photon that is observed from impact parameter $\alpha\equiv r/d_{\rm A}(z)$ (with $d_{\rm A}(z)$ being the angular diameter distance to redshift $z$) contributes $\cos^2 \chi$ to $I_l$ and $\sin^2 \chi$ to $I_r$ for Rayleigh (or wing) scattering \citep{RL99}. 

However, for resonant scattering a photon contributes $p_{\rm R} \cos^2 \chi+\frac{1}{2}(1-p_{\rm R})$ to $I_l$ and $p_{\rm R} \sin^2 \chi+\frac{1}{2}(1-p_{\rm R})$ to $I_r$. Here, $p_{\rm R}=({1+\cos^2\theta})/(\frac{11}{3}+\cos^2\theta)$ is the probability that a core photon was Rayleigh scattered in
the last scattering event (see Appendix A1), in which $\theta$ is the angle between the propagation directions of the incoming and outgoing photons at last scattering ($\cos \theta={\bf n \cdot n'}$). The fact that $p_{\rm R}$ depends on angle $\theta$ has a simple reason.  Although resonant scattering may be viewed as a superposition of Rayleigh scattering and isotropic scattering of weights $1/3$ and $2/3$, these weights are averaged over solid angle. The actual weight of each process is angle dependent (see Appendix~\ref{app:super}). In other words, in the case of resonant scattering the intensities in radial and azimuthal directions are obtained by scaling down the intensities obtained from Rayleigh scattering by a factor of $p_{\rm R}$, and by adding an unpolarized intensity of magnitude $1-p_{\rm R}$.  

The observed fractional polarization at impact parameter $\alpha$ is given by
\begin{equation}
\mathcal{P}(\alpha)=\frac{|I_l(\alpha)-I_r(\alpha)|}
{I_l(\alpha)+I_r(\alpha)}.
\end{equation} 
The fractional polarization can also be expressed in terms of the Stokes
parameters $Q$ and $I$ as $\mathcal{P}=|Q|/I$. In this paper, we always
find that $I_{\rm l}<I_{\rm r}$, and $Q<0$. This may be represented by
arrows drawn in concentric rings around the central Ly$\alpha$ emitter,
where the size of an arrow indicates the magnitude of $\mathcal{P}$ (see
Fig~\ref{fig:scheme2}).

\begin{figure*}
\vbox{\centerline{\epsfig{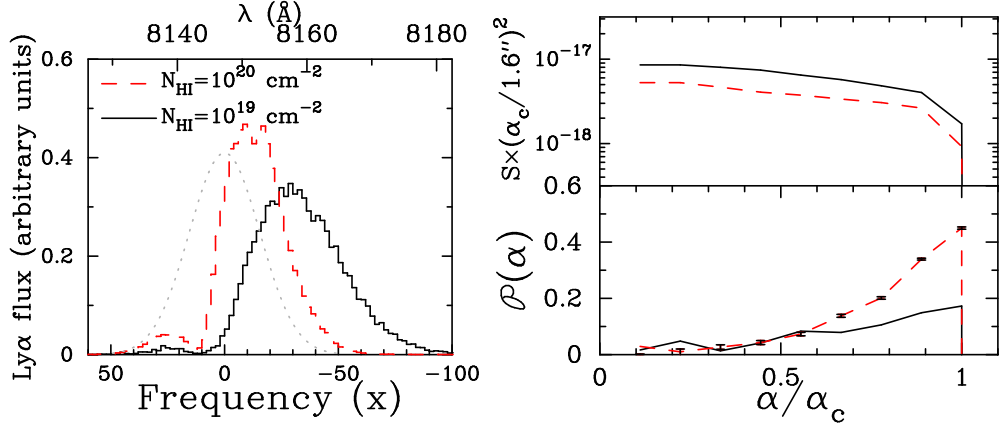}}}
\caption[]{The observable properties of the Ly$\alpha$ radiation emitted by
a galaxy which is surrounded by a single thin spherical shell of gas that expands at a speed $v_{\rm exp}=200$ km s$^{-1}$ and contains neutral hydrogen with a
column density of either $N_{\rm HI}=10^{19}$ cm$^{-2}$ ({\it red dashed lines})
or $N_{\rm HI}=10^{20}$ cm$^{-2}$ ({\it solid lines}). These parameter
choices represent the generic conditions in galaxies surrounded by a
superwind driven outflow (see text). The {\it dotted line} in the {\it left
panel} shows the Gaussian Ly$\alpha$ emission line prior to scattering. The
figure shows that backscattering off the expanding outflow results in a
systematic redshift of the Ly$\alpha$ line. The {\it top and bottom right
panels} show the surface brightness profile (units are erg s$^{-1}$ cm$^{-2}$ arcsec$^{-2}$) and fractional polarization
$\mathcal{P}(\alpha)$ as a function of $\alpha$, respectively. Here, the
amplitude of $\alpha$ and the surface brightness $S(\alpha)$ are written in
terms of $\alpha_{\rm c}$, which is related to the distance of the
expanding shell from the galaxy (see text). The Figure shows that the
polarization reaches values as high as $\sim 40\%$.}
\label{fig:back}
\end{figure*}

\item Monte-Carlo calculations of Ly$\alpha$ transfer can be accelerated by
skipping scattering events that occur in the line core \citep[e.g.][]{Ahn02,D06}. One has to be somewhat careful when applying this technique in calculations of polarization, as it reduces the number of scattering events that a Ly$\alpha$ photon encounters and hence distorts its resulting polarization state. However, we have verified explicitly that the polarization properties of Ly$\alpha$ emerging from optically thick clouds are correctly calculated by the accelerated scheme. Generally speaking, the accelerated Monte-Carlo scheme yields the correct polarization whenever this scheme reliably computes other observable quantities such as the spectrum and surface brightness profile.

\end{itemize}

Lastly, we point out that the expressions for $p(\theta)$ and $\Pi(\theta)$ in
\S~\ref{sec:RT2} were derived under the assumption that the radiation field
prior to the scattering event was unpolarized. The precise functional forms
of $p(\theta)$ and $\Pi(\theta)$ are different when the radiation field
prior to scattering is polarized (see Appendix A1). The procedure of propagating the polarization vector in a single Rayleigh scattering event that was outlined above, naturally accounts for the polarization dependence of $\Pi(\theta)$ and $p(\theta)$ (see Appendix A3). For resonantly scattered Ly$\alpha$ the situation is more complex and the density matrix formalism should be used \citep[e.g.][]{Lee94,Lee98,Ahn02}. However, the precise dependence of $p(\theta)$ and $\Pi(\theta)$ on initial polarization does not affect the calculations regarding resonant scattering. The maximum fractional polarization we find for resonantly scattered Ly$\alpha$ is $7\%$ (see \S~\ref{sec:igm}). The typical polarization of the Ly$\alpha$ radiation field 'seen' by atoms involved in the scattering process is even lower. This only introduces changes in the phase function at the level of a
few per cent, and we have verified that our results regarding resonantly
scattered Ly$\alpha$ are insensitive to the precise form of the phase
function.

\section{Ly$\alpha$ Polarization Around High-Redshift Sources}
\label{sec:result}

Next we calculate the expected polarization of the Ly$\alpha$ line in a set
of models which span the likely environments of high-redshift
galaxies. Unless otherwise stated, we assume the redshift of Ly$\alpha$
sources to be $z=5.7$ in all cases. The Monte-Carlo radiative transfer
calculations rely on stacking individual photons into bins, and the
errorbars shown in some figures were calculated assuming Poisson
fluctuations in the number of photons within a given bin.

\subsection{Backscattering off Galactic Superwinds}
\label{sec:wind}

\subsubsection{Description of the Model}

A major fraction of Lyman break galaxies (LBGs) at high redshifts show
evidence of being surrounded by outflowing enriched gas
\citep{Shapley03}. Scattering of Ly$\alpha$ photons by neutral hydrogen
atoms in these outflows may cause the observed Ly$\alpha$ line to be
redshifted systematically relative to the systemic velocity of the galaxy
\citep{Ahn03}. This redshift is attributed to the Doppler boost that
Ly$\alpha$ photons experience when they scatter off the outflow on the far
side of the galaxy back towards the observer (hence the term
'backscattering'). This redshifted Ly$\alpha$ flux is less prone to
resonant scattering in the IGM, and is therefore more easily observable
than Ly$\alpha$ photons that were not backscattered.

\begin{figure}
\vbox{\centerline{\epsfig{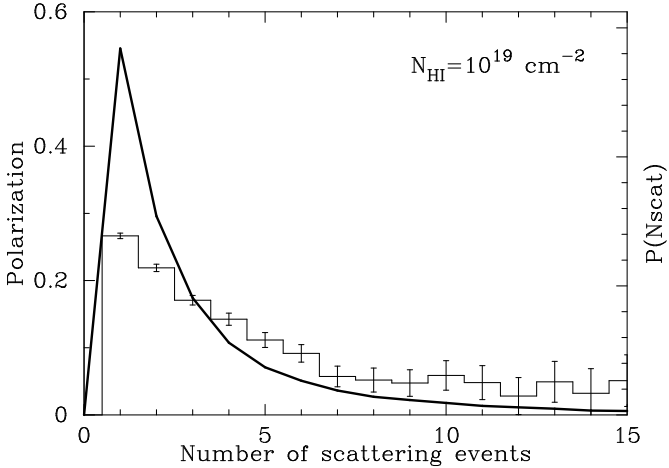}}}
\caption[]{The {\it histogram} shows the angle-averaged polarization as a
function of the number of times Ly$\alpha$ photons scatter for the superwind model with $N_{\rm HI}=10^{19}$ cm$^{-2}$. The angle-averaged polarization is largest for photons that scatter only once ($\langle P(\alpha)\rangle\sim 30\%$), and almost vanishes for $N_{\rm scat}\geq 10$. The {\it thick solid line} shows the probability distribution
of the number of times Ly$\alpha$ photons scatter. Clearly, the majority of
photons scatter only once, which causes the observed Ly$\alpha$ flux to be
highly polarized.}
\label{fig:nscat}
\end{figure}

\begin{figure}
\vbox{\centerline{\epsfig{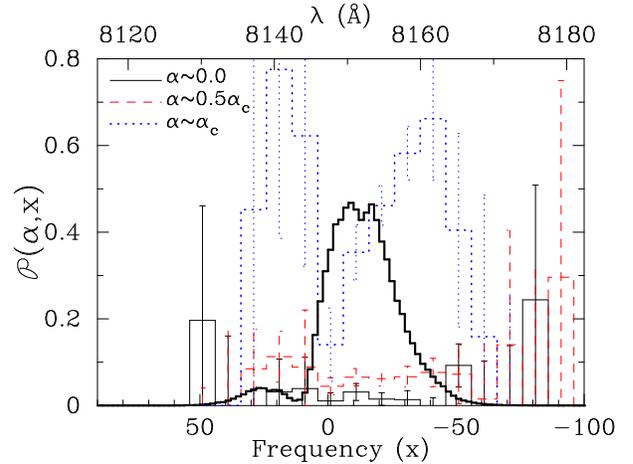}}}
\caption[]{The frequency dependence of the polarization at $\alpha=0$ ({\it
thin black histogram}), $0.5\alpha_{\rm c}$ ({\it red dashed histogram}),
and $\alpha_{\rm c}$ ({\it blue dotted histogram}). The polarization
increases with impact parameter. The {\it thick solid line} shows the
observed spectrum which was also shown in Fig~\ref{fig:back} (the units are arbitrary). The bulk of
the photons have frequencies $-50 <x<10$. Within this frequency
range, the polarization increases towards the red. This is because the
redder Ly$\alpha$ photons appear farther from resonance in the frame of the
expanding gas; consequently they scatter less and achieve a higher
polarization amplitude (see Fig~\ref{fig:nscat}). This generic frequency
dependence of the polarization amplitude can be used as a fingerprint of
outflows.  The frequency dependence is reversed for infalling gas (see
\S~\ref{sec:cool}).}
\label{fig:impact}
\end{figure}

The impact of the outflow on the observable properties of the Ly$\alpha$
line depends on various parameters including the outflow speed, $v_{\rm
exp}$, the total column density of neutral hydrogen atoms in the outflow,
$N_{\rm HI}$, and the dust content and distribution inside the outflow
\citep{Hansen06,Verhamme06}. We show results for a single expanding thin shell of gas with $N_{\rm HI}=10^{19}$ cm$^{-2}$ and $N_{\rm HI}=10^{20}$ cm$^{-2}$, based on column densities that are observed in Ly$\alpha$ emitting galaxies \citep{Kunth98,Verhamme07}. We discuss the choice of our model in more detail after presenting our results below. The outflow speed is typically of order the circular velocity of the host dark matter halo, $v_{\rm circ}$ \citep[e.g.][]{Fur}. We assume the galaxy to be embedded in a halo of total mass $M_{\rm tot}=3\times 10^{11} M_{\odot}$, which corresponds to $v_{\rm circ}=200$ km s$^{-1}$.

\subsubsection{Results}

The emitted Ly$\alpha$ spectrum prior to scattering is assumed to be a
Gaussian with a velocity width $\sigma=v_{\rm circ}$ \citep{igm}
and is denoted by the {\it dotted line} in the {\it left panel} of
Figure~\ref{fig:back}. The {\it dashed (solid) lines} represent the
observed spectrum for an outflow with $N_{\rm HI}=10^{19}$ cm$^{-2}$
($10^{20}$ cm$^{-2}$).  The Figure shows that backscattering off the
expanding outflow results in a systematic redshift of the Ly$\alpha$ line
(as mentioned above), with an increase in redshift as $N_{\rm HI}$
increases.  The {\it top right panel} shows that the surface brightness
profile remains flat out to some value of $\alpha=\alpha_{\rm c}$ after
which it drops to zero (as there is no scattering outside the shell). The
magnitude of $\alpha_{\rm c}$ and the surface brightness $S(\alpha)\equiv
I_{\rm l}(\alpha)+I_{\rm r}(\alpha)$ 
are determined by the radius of the expanding shell. For example, if the
outflow is located at a distance $d=10$ kpc from the central galaxy (and
its thickness $\ll d$), then the surface brightness profile drops to zero
at $\alpha_{\rm c}=1.6'' (d/10\hs {\rm kpc})$ for a galaxy at $z=5.7$. The
mean surface brightness scales approximately\footnote{Fig~\ref{fig:back} shows that the surface brightness remains almost constant out to $\alpha/\alpha_{\rm c}\sim0.8$. The mean surface brightness, $\langle S \rangle$, was therefore estimated by dividing the total scattered flux by $\pi\alpha_{\rm c}^2$.} as $\langle S \rangle \approx \mathcal{X} \times 10^{-18}\hs (\alpha_{\rm c}/1.6'')^{-2}(L_{{\rm Ly}\alpha}/10^{43}\hs {\rm erg}\hs{\rm s}^{-1})$ erg s$^{-1}$ cm$^{-2}$ arcsec$^{-2}$, where $\mathcal{X}=3$ for $N_{\rm HI}=10^{19}$ cm$^{-2}$, and $\mathcal{X}=5$ for $N_{\rm HI}=10^{20}$cm$^{-2}$.

The {\it bottom right panel} shows that the fractional polarization,
$\mathcal{P}(\alpha)$, increases towards larger values of $\alpha$ and
reaches a maximum near $\alpha_{\rm c}$. The polarization amplitude obtains
values as high as $\sim 40\%$ for $N_{\rm HI}=10^{19}$ cm$^{-2}$, and $\sim
18\%$ for $N_{\rm HI}=10^{20}$ cm$^{-2}$.

The reason for the overall large values of the polarization is easy to
understand. Although the line center optical depth through the outflow is
very large for both models, $\tau_{0}=5.9\times 10^5(N_{\rm HI}/10^{19}\hs
{\rm cm}^{-2})$ (see Eq~\ref{eq:tau0}), most photons reside in the wings of
the line profile when they reach the expanding shell. For example, in the
model with $N_{\rm HI}=10^{19}$ cm$^{-2}$, the optical depth is less than
unity for $63\%$ of the photons when they enter the outflow for the first
time. Therefore, most photons scatter only once in the wing of the line
profile before escaping to the observer.

The above interpretation is quantified by Figure~\ref{fig:nscat}, in which
the {\it thick solid line} shows the probability distribution for the
number of scattering events encountered by the Ly$\alpha$ photons,
$P(N_{\rm scat})$. The function $P(N_{\rm scat})$ peaks at $N_{\rm scat}=1$
and rapidly decreases for increasing $N_{\rm scat}$. Also shown as the {\it
histogram} is the angle-averaged polarization of the photons as a function
of $N_{\rm scat}$. The polarization is largest for photons that scatter
only once ($\langle P(\alpha)\rangle\sim 30\%$), and nearly vanishes for
$N_{\rm scat}\geq 10$. The decrease in the polarization amplitude with
increasing $N_{\rm scat}$ reflects that the radiation field becomes
increasingly isotropic as the number of scattering events increases. As
$N_{\rm HI}$ increases, the fraction of photons that escape after a single
scatter decreases, and the overall polarization declines.

Figure~\ref{fig:impact} shows the frequency dependence of the
polarization at three different impact parameters, namely $\alpha=0$ ({\it
thin black histogram}), $0.5\alpha_{\rm c}$ ({\it red dashed histogram}),
and $\alpha_{\rm c}$ ({\it blue dotted histogram}).  As already illustrated
in Figure~\ref{fig:back}, the polarization increases with increasing impact
parameter. The {\it thick solid line} shows the observed spectrum which was
previously shown in Fig~\ref{fig:back}. The bulk of photons have
frequencies $-50 <x<10$. Within this frequency range, the
polarization increases towards the red. This is because the redder
Ly$\alpha$ photons appear farther from resonance in the frame of the gas;
they therefore scatter less and achieve a higher polarization amplitude
(see Fig~\ref{fig:nscat}). This frequency dependence of the polarization
can be used as a fingerprint to distinguish outflows from infall (see
\S~\ref{sec:cool}). 


One may wonder whether the above results depend sensitively on the assumed model: e.g. do the result change for a continuous, dusty, wind with a radial dependence of its outflow speed? A wind for which the outflow speed increases linearly with radius would resemble the IGM with a Hubble flow as discussed in \citet{RL99}, who found even higher levels of polarization. Furthermore, scattering through an optically thick {\it collapsing} gas cloud also results in a comparable polarization ($\mathcal{P}_{\rm max} \sim 35\%$, see \S~\ref{sec:cool}). Changing the sign of the velocity of the gas only affects the frequency dependence of the polarization (see \S~\ref{sec:cool}). The column density of neutral hydrogen in both these models is $N_{\rm HI}\gg 10^{20}$ cm$^{-2}$, which implies that polarized Ly$\alpha$ is also expected for outflow models with larger column densities and non-zero velocity gradients. 

Furthermore, the presence of dust in the expanding wind can boost the polarization. The reason is that photons that scatter only once obtain the highest level of polarization, while it decreases for photons that scatter multiple times (Fig~\ref{fig:nscat}). However, photons that scatter multiple times traverse a longer path through the wind, which enhances the probability for absorption by dust. Hence, dust can preferentially quench the low polarization photons, which would strengthen our results.

\begin{figure*}
\vbox{\centerline{\epsfig{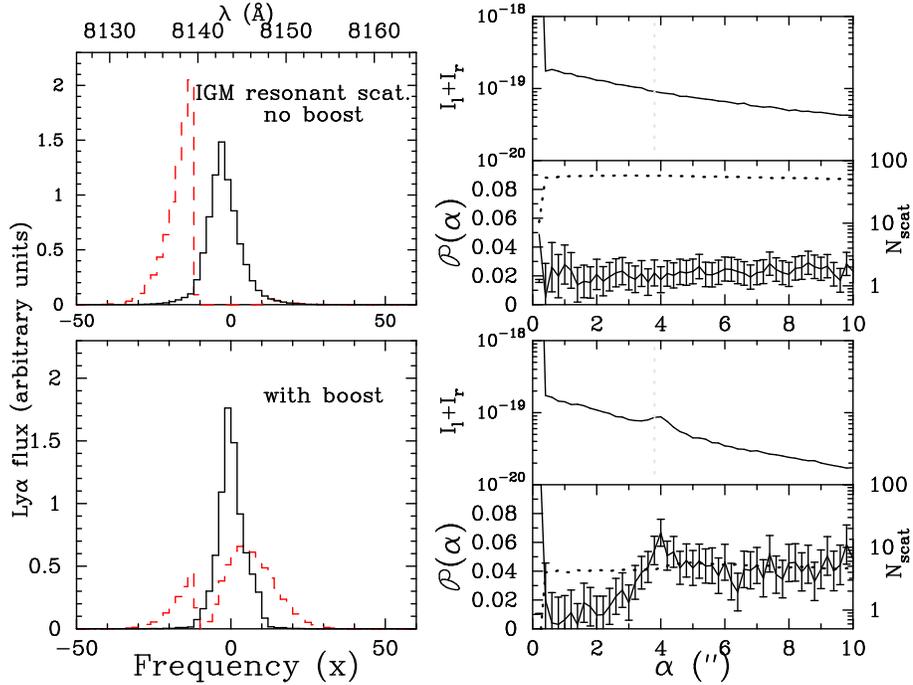}}}
\caption[]{The observable properties of Ly$\alpha$ radiation that is
resonantly scattered in the intergalactic medium (IGM). The cases
considered here represent galaxies that are not surrounded by a
superwind-driven outflow and complement the examples shown in
Fig~\ref{fig:back} (see text). The IGM is assumed to have a peculiar infall velocity (see text). The {\it upper left panel} shows the spectrum of Ly$\alpha$ radiation that was transmitted without scattering once as the {\it red dashed line}. This spectrum reflects the line-emission spectrum observed from the
galaxy (see text). The spectrum of resonantly scattered Ly$\alpha$ radiation is shown as the {\it solid line} (both spectra are normalized).  The {\it upper right panels} show that the fractional polarization is only $\mathcal{P}(\alpha)\sim 2\%$. The main reason is that
each photon resonantly scatters $\sim 50$ times, as shown by the {\it
dotted line}.  The {\it lower panels} show the results for the same
calculation but with an increased level of the ionizing background. The
boost in the ionizing background increases the transmission on the
blue-side of the line ({\it lower left panel}). In this case each
Ly$\alpha$ photon scatters on average only a few times, and the scattered
Ly$\alpha$ is polarized up to $\sim 7\%$ ({\it lower right panels}). Thus,
the polarization of resonantly scattered Ly$\alpha$ photons in the IGM is
low compared to that in \S~\ref{sec:wind}, and its level depends on the
local level of the ionizing background.}
\label{fig:igm1}
\end{figure*}
\subsection{Resonant Scattering in the IGM}
\label{sec:igm}

\subsubsection{Description of the Model}

Even after reionization, residual neutral hydrogen in the IGM can scatter
up to $\gsim 90\%$ of the Ly$\alpha$ emitted by galaxies out of our line of
sight \citep{Santos04,igm}. This scattered Ly$\alpha$ radiation would
appear as a faint Ly$\alpha$ halo surrounding the central Ly$\alpha$
emitting galaxy \citep{Loeb99}. We note that back-scattered Ly$\alpha$
radiation is much less prone to resonant scattering in the IGM.  Therefore
we now consider galaxies that are not surrounded by wind-driven outflows,
and examine examples that complement the cases discussed previously.

We adopt the fiducial model described by \citet{igm} to examine the impact
of the IGM on the observed Ly$\alpha$ line profile of a $z=5.7$ galaxy
embedded in a dark matter halo with a total mass of $M_{\rm
tot}=10^{11}M_{\odot}$. The halo has a virial radius $r_{\rm vir}=23$ kpc
and a circular velocity $v_{\rm circ}=133$ km s$^{-1}$. The galaxy is
assumed to form stars at a rate of $\dot{M}_*=10M_{\odot}~{\rm yr}^{-1}$,
and the escape fraction of its ionizing photons is taken to be $f_{\rm
esc}=0.1$. The resulting Ly$\alpha$ luminosity of the galaxy is
\llya$=2\times 10^{43}$ erg s$^{-1}$ for a gas metallicity of
$Z=0.05Z_{\odot}$. Gas within the virial radius is assumed to be fully ionized ($x_{\rm H}=0$) and no scattering occurs here. Just outside the virial radius the IGM density, $\rho(r_{\rm vir})=20\bar{\rho}$, where $\bar{\rho}$ is the mean baryon density of the Universe. The density decreases as $\rho(r)\propto r^{-1}$ until it approaches $\bar{\rho}$. Just outside the virial radius, the IGM is assumed to be collapsing at $v_{\rm IGM}\sim v_{\rm circ}$. Only at sufficiently large radii ($r>10 r_{\rm vir}$) the IGM is comoving with the Hubble flow (see Barkana 2004 and Dijkstra et al 2007b for a more detailed description and a motivation of this model)\footnote{\citet{igm} account for gas clumping in the IGM. Most of the
volume of the IGM consists of interclump gas that is underdense. For this
reason the mean IGM opacity to Ly$\alpha$ photons is lower in a clumpy IGM
at a fixed value of the ionizing background (Dijkstra et al 2007b). Because
clumps are not included in the radiative transfer code, we increased the
ionizing background (compared to the clumpy case) in order to match the
observed mean transmission of the IGM at a redshift $z$, $\langle
e^{-\tau}\rangle(z)$ \citep[e.g.][]{Fan06}.}.

We also examine a case where we artificially boost the photoionization rate
in the IGM. This boost may represent: ({\it i}) an enhanced local ionizing
background due to clustering of undetected surrounding sources
\citep{WL05,igm}; ({\it ii}) an enhanced ionizing background due to the
presence of a nearby quasar; ({\it iii}) a galaxy at a lower redshift where
the residual neutral hydrogen fraction in the IGM is lower; ({\it iv})
an enhanced local photoionization rate due to vigorous star formation in
the galaxy itself, or due to an enhanced escape fraction of ionizing
photons; ({\it v}) a lower neutral fraction because of a lower overall density of hydrogen nuclei along the line of sight (e.g. for a galaxy on the edge of a void.)


\subsubsection{Results} 

In the {\it top left panel} of Figure~\ref{fig:igm1} we show the spectrum
of Ly$\alpha$ radiation that was transmitted without scattering once ({\it dashed line}), which represents the Ly$\alpha$ spectrum observed from the galaxy. This spectrum was the focus of the analysis of e.g Santos (2004) and \citet{igm}. The plot shows that the blue side of the Ly$\alpha$ line observed from the galaxy ($x>0$) is eliminated by the IGM. The suppression extends into the red part of the line because of resonant scattering by residual neutral hydrogen gas that is falling onto the galaxy. Therefore, the peak of the observed spectrum is redshifted relative to the true line center by an amount which is set by the gas infall velocity. The redshift in velocity is $v_{\rm IGM}(r_{\rm vir})=-v_{\rm circ}$. The spectrum of the resonantly scattered Ly$\alpha$ halo is shown as the {\it solid line}. 

\begin{figure*}
\vbox{\centerline{\epsfig{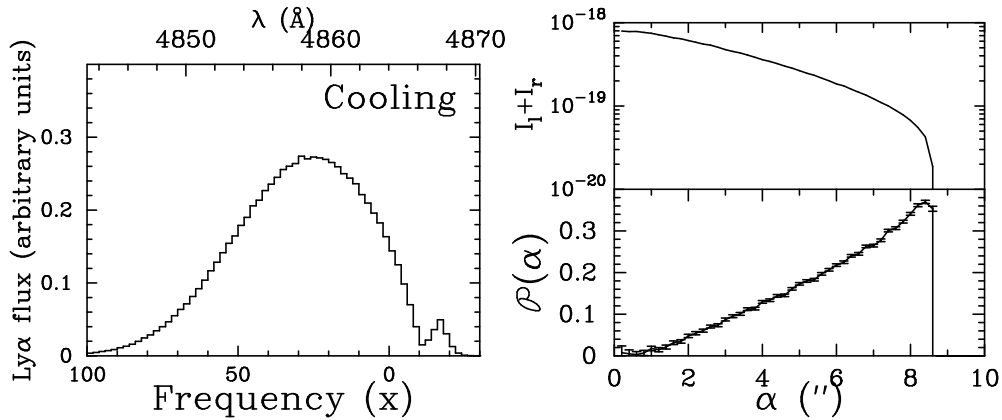}}}
\caption[]{The observable properties of Ly$\alpha$ radiation emerging from
a neutral collapsing gas cloud. The model represents protogalaxies in the
process of their assembly. The {\it left panel} shows that the spectrum
emerging from such a cloud has a systematic blueshift. The {\it right
panel} shows that the Ly$\alpha$ surface brightness profile (units are erg s$^{-1}$ cm$^{-2}$ arcsec$^{-2}$) is rather flat
({\it top}), and that it is highly polarized. The fractional polarization
increases roughly linearly towards the edge of the cloud, and reaches a
maximum value of $\mathcal{P}_{\rm max}\sim 35\%$. }
\label{fig:cool}
\end{figure*}

The {\it top right panel} of Figure~\ref{fig:igm1} shows the surface
brightness profile of the scattered Ly$\alpha$ halo as a function of $\alpha$. The observed surface brightness of the scattered Ly$\alpha$ photons is $\sim 10^{-19}$ erg
s$^{-1}$ cm$^{-2}$ arcsec$^{-2}$ near the center, and decreases with
increasing angular separation from the galaxy. Overall, the fractional
polarization is small, $\mathcal{P}(\alpha)\sim 2\%$.  Besides the fact
that resonantly scattered Ly$\alpha$ radiation is always less polarized
than the radiation in the line wings (see \S~\ref{sec:RT2}), there is the
fact that when the Ly$\alpha$ enters resonance, it typically scatters
multiple times before escaping towards the observer. This is shown by the
{\it dotted line} which gives the mean number of scattering events for the
Ly$\alpha$ photons, $\langle N_{\rm scat}\rangle$, as a function of impact
parameter $\alpha$. In this model $\langle N_{\rm scat}\rangle\sim
50$. Hence, the Ly$\alpha$ radiation field of resonantly scattered
Ly$\alpha$ photons will locally be close to isotropic, resulting in a low polarization.

The {\it lower panels} in Figure~\ref{fig:igm1} show the results for the model in which the ionizing background was boosted. The {\it lower left panel} shows that boosting the photoionization rate decreases the fraction of residual neutral hydrogen gas, which in turn increases the transmission on the blue-side of the line. The {\it lower right panels} show that the surface brightness profile falls slightly faster with a slight peak in the surface brightness profile near the virial radius. The reason for this peak is that in our model the IGM is densest right outside the virial radius, which results in the largest effective optical depth here (this also produces the dip in the spectrum). In this model each Ly$\alpha$ photon scatters on average only a few times which causes the scattered Ly$\alpha$ to be polarized up to $\sim 7\%$. The mean polarization decreases at $\alpha<\alpha_{\rm vir}=4''$. This is because in our model no scattering occurs inside the virial radius, which extends out to $\alpha_{\rm vir}$. Therefore, the relative number of photons that was scattered at almost-right-angles decreases at $\alpha<\alpha_{\rm vir}$, which in turn decreases the polarization. For example, scattered photons emerging from $\alpha=0''$ were all scattered by $\theta=180^{\circ}$ and are therefore unpolarized ($\Pi(180^{\circ})=0$). 

Our radiative transfer code does not account for gas clumping in the IGM (in our simulation the gas density is a smooth function of radius $r$). Gas clumping affects the predicted polarization: Ly$\alpha$ photons that resonate with gas inside denser clumps scatter more than calculated in our model, and are therefore polarized at lower levels (see Fig~\ref{fig:nscat}). On the other hand, photons that resonate with gas in the interclump medium scatter less and are expected to me more strongly polarized. Since this interclump medium has the largest volume filling factor, gas clumping results in net higher level of polarization. The impact of gas clumping on the predicted polarization properties - at a fixed level of the ionizing background - can therefore be mimicked by increasing the level of the ionizing background somewhat. 

In summary, the polarization of resonantly scattered Ly$\alpha$ photons in
the IGM is low compared to the values derived in \S~\ref{sec:wind}, with
the maximum polarization level reaching values of $\mathcal{P}_{\rm max}
\lsim 7\%$. The polarization amplitude depends strongly on the local level
of the ionizing background. This is because the polarization decreases with
increasing effective optical depth, $\tau_{\rm eff}$, which in turn
increases with a decreasing local photoionization rate. 

\subsection{Scattering in Optically Thick Clouds}
\label{sec:cool}
\subsubsection{Description of the Model}

\citet{D06} performed Monte-Carlo calculations of Ly$\alpha$ radiative
transfer through optically thick, spherically symmetric, collapsing gas
clouds, which represent simplified models of protogalaxies in the process
of their assembly. Here we compute the Ly$\alpha$ polarization properties
for one of the fiducial models presented by \citet{D06}, in which a dark
matter halo of mass $M_{\rm tot}=5.2 \times 10^{11}M_{\odot}$ collapses at
$z=3$ while it continuously emits Ly$\alpha$ over a spatially extended
region (see 'model {\bf 1.}' of Dijkstra et al 2006).

\subsubsection{Results}
\begin{figure}
\vbox{\centerline{\epsfig{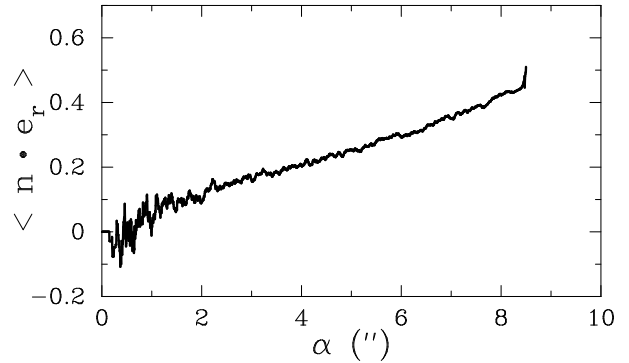}}}
\caption[]{The level of anisotropy of the Ly$\alpha$ radiation field, quantified by $\langle {\bf n} \cdot {\bf e}_{\rm r}\rangle$ (see text), is shown. For an isotropic radiation field $\langle {\bf n} \cdot {\bf e}_{\rm r}\rangle=0$. Clearly, $\langle{\bf n} \cdot {\bf e}_{\rm r}\rangle >0$ and increases with radius. Photons that are scattered towards the observer were therefore preferentially moving radially outward prior to their last scattering event. This Figure shows clearly that the Ly$\alpha$ radiation field becomes increasingly anisotropic with radius, which results in an increasingly large fractional polarization with impact parameter (Fig~\ref{fig:cool}).}
\label{fig:dir}
\end{figure}
The {\it left panel} of Figure~\ref{fig:cool} shows that the Ly$\alpha$
emerges with a systematic blueshift, which is due to energy transfer from
the gas to the photons, combined with a reduced escape probability for
photons in the red wing of the line profile \citep[see][for a detailed
discussion]{D06}. 
The {\it right panels} of
Figure~\ref{fig:cool} show the surface brightness profile ({\it top}), and
the fractional polarization $\mathcal{P}(\alpha)$ ({\it bottom}) as a
function of $\alpha$. The fractional polarization again reaches values as
high as $\mathcal{P}_{\rm max}=35\%$. This may be surprising given the fact
that each photon scatters numerous times before emerging from the cloud. As
mentioned in \S~\ref{sec:RT1}, Ly$\alpha$ photons generally escape from an
optically thick medium in the line wing, where they acquire larger
polarizations than in the line core. Furthermore, each spherical shell
inside the cloud must see a net outward flow of Ly$\alpha$ photons, or
 else the cloud would not cool. 

This is quantified in Figure~\ref{fig:dir} which shows the radial dependence of the average value of ${\bf n} \cdot {\bf e}_{\rm r}$. Here ${\bf n}$ denotes a photon's propagation direction prior to the last scattering event (as in Fig~\ref{fig:scheme}), and ${\bf e}_{\rm r}$ denotes a unit-vector that is pointing radially outward from the center of the cloud. In this convention ${\bf n} \cdot {\bf e}_{\rm r}=1$ (${\bf n} \cdot {\bf e}_{\rm r}=-1$) for a photon that is moving radially outward (inward). For an isotropic radiation field $\langle {\bf n} \cdot {\bf e}_{\rm r}\rangle=0$. Clearly,   $\langle {\bf n} \cdot {\bf e}_{\rm r}\rangle$ is positive at all radii, and increases with radius. In other words, photons that are scattered towards the observer were preferentially moving radially outward prior to the scattering event. Furthermore, this anisotropy in the radiation field\footnote{We point out that the actual Ly$\alpha$ radiation field that is 'seen' by each atom in the cloud is dominated by photons that are resonantly scattering, and is almost completely isotropic. However, these photons have a negligible probability of escaping towards the observer and do not affect the observable Ly$\alpha$ polarization properties of the cloud.} increases with radius, which results in an increasingly large fractional polarization with impact parameter (Fig~\ref{fig:cool}).

\begin{figure}
\vbox{\centerline{\epsfig{file=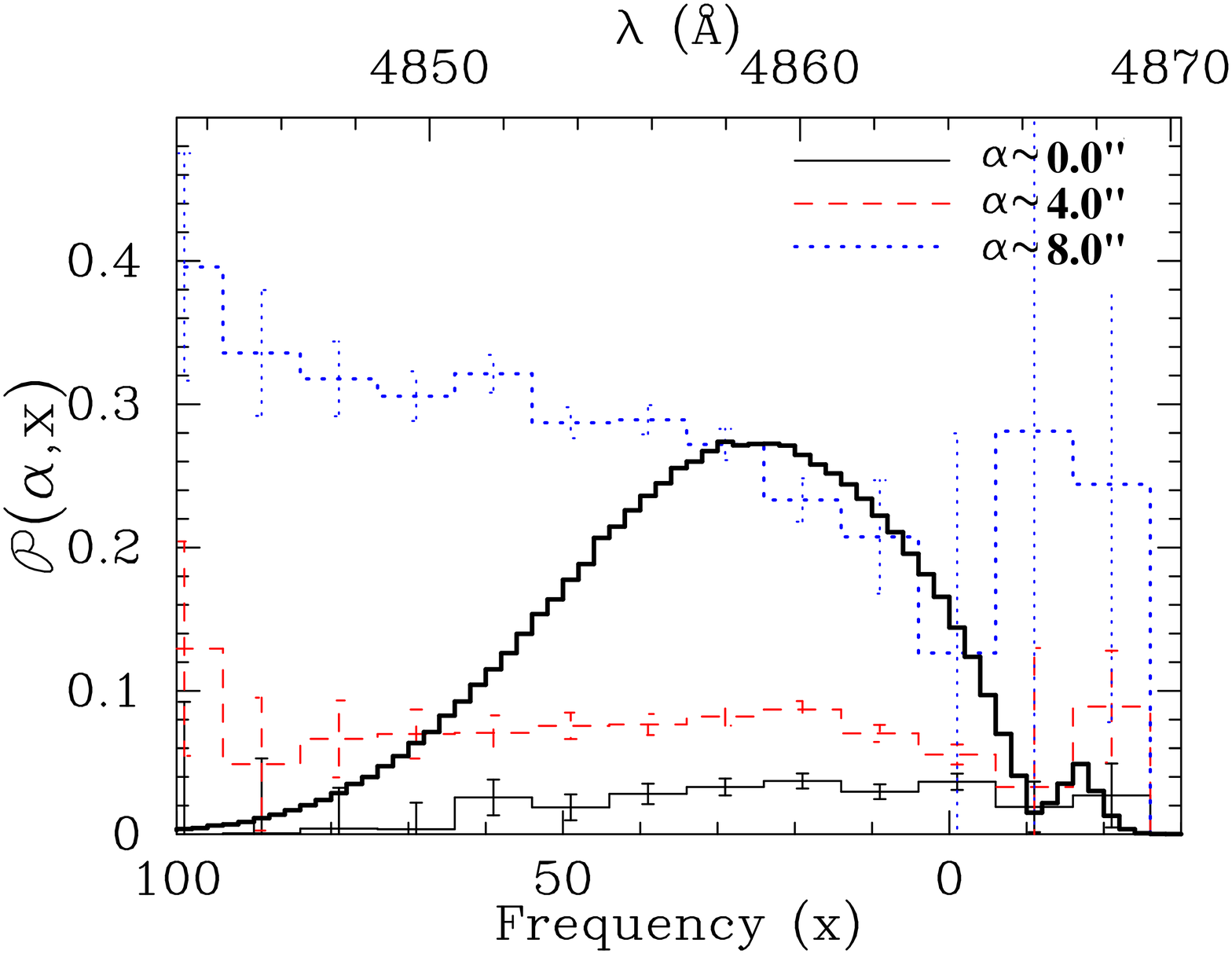,angle=0,width=7.5truecm}}}
\caption[]{The frequency dependence of the polarization at three different
angular separations from the galaxy: $\alpha\sim 0$, $4.0$, and $8.0$
arcseconds. The polarization increases with separation (as was already
demonstrated in Fig~\ref{fig:cool}). The {\it thick solid line} shows the
spectrum (arbitrary units, previously shown in Figure~\ref{fig:cool}). The bulk of the photons emerge with frequencies $-10<x<80$. Within this frequency range, the polarization increases towards higher frequencies, in sharp contrast
with the outflow model discussed in \S~\ref{sec:wind}.}
\label{fig:impactcool}
\end{figure}

In Figure~\ref{fig:impactcool} we show the frequency dependence of the
polarization at 3 different angular separations: $\alpha\sim 0$, $4.0$, and
$8.0$ arcseconds. The {\it thick solid line} is the spectrum shown
previously in Figure~\ref{fig:cool}, indicating that the bulk of the
photons emerge with frequencies $-10<x<80$. Within this frequency range,
the polarization increases towards the blue. The reason for this wavelength
dependent polarization is that Ly$\alpha$ photons that are far in the blue
wing of the line profile, appear even farther in the wing in the frame of
the infalling gas. Therefore, the escape probability of Ly$\alpha$ photons
increases, and the radiation field becomes increasingly anisotropic towards higher photon frequencies. This in turn increases the polarization towards higher frequencies. Lastly, it should be pointed out that the frequency dependence of the polarization applies mostly to the case $\alpha\sim 8.0$ arcseconds, and similarly to the case $\alpha \sim \alpha_c$ that was shown in Figure~\ref{fig:impact}. Hence, the frequency dependence of the polarization is evident only where the overall (frequency averaged) polarization is significant ($\mathcal{P}\gsim 10\%$).

\section{Discussion}
\label{sec:discuss}
\subsection{Constraints on EoR and High-z Galaxy Formation}

The Ly$\alpha$ emission needs to be spatially resolved in order for its
polarization to be measured. Over the past decade, a large number of
spatially extended Ly$\alpha$ emitters, also known as Ly$\alpha$ 'blobs',
were discovered \citep{Steidel00,Matsuda04,Saito06}. In fact, the size
distribution of Ly$\alpha$ sources is continuous and Ly$\alpha$ blobs
simply represent the largest and rarest objects in this distribution
\citep{Matsuda04}. This implies that there is no shortage of spatially
extended Ly$\alpha$ sources in the Universe.

As was shown in \S~\ref{sec:result}, the polarization properties of
spatially resolved Ly$\alpha$ emission encode information about the medium
through which the Ly$\alpha$ photons propagate, especially in combination
with spectral information (Fig~\ref{fig:impact} and
Fig~\ref{fig:impactcool}). For example, large fractional polarizations
(tens of percent) are expected for Rayleigh scattering. An increase in the
polarization towards longer wavelengths would be indicative of
backscattering off an expanding outflow. Such an inference could be
particularly important because ({\it i}) it provides evidence that the
extended Ly$\alpha$ emission results from scattering ({\it ii}) it allows
to better gauge the impact of resonant scattering in the IGM on the
observed Ly$\alpha$ line profile, since back-scattered Ly$\alpha$ radiation
is much less prone to resonant scattering in the IGM. Understanding the
impact of resonant scattering on the Ly$\alpha$ line profiles is essential
for ({\it iia}) deriving constraints on the ionization state of the IGM
\citep{LF}, and ({\it iib}) interpreting  Ly$\alpha$ emitters
with large equivalent widths \citep{MR02,DW07}.

In this paper we have shown numerical examples for the polarization
properties of Ly$\alpha$ radiation that is either {\it (i)} resonantly
scattered in the IGM; {\it (ii)} backscattered in an outflow; or {\it
(iii)} associated with cooling radiation from a collapsing
protogalaxy. There are other classes of spatially extended Ly$\alpha$
sources that have not been discussed:

\begin{itemize}

\item Gas in close proximity to a bright source of ionizing radiation
(e.g. a quasar) may 'glow' in fluorescent Ly$\alpha$ emission
\citep{HR01,Weidinger04}. If the recombining Ly$\alpha$ emitting gas is sufficiently close to the quasar or the quasar is particularly bright, then this gas is optically thin to Ly$\alpha$ photons and the emergent Ly$\alpha$ radiation is not expected to be polarized. However, fluorescent Ly$\alpha$ emission is also expected from Lyman limit systems and Damped Ly$\alpha$ systems \citep[][]{Gould96,Cantalupo05,Adelberger06,Cantalupo07}, which are optically thick to Ly$\alpha$. Especially in cases where the gas is illuminated anisotropically, scattering could lead to a detectable polarization. 


\item Spatially extended Ly$\alpha$ emission may be generated by gas that
was shock heated by a superwind-driven outflow
\citep[e.g.][]{Taniguchi00,Mori04}. The backscattering mechanism described
in \S~\ref{sec:wind} may also operate here, and this could lead to a
measurable polarization amplitude.

\item Extended luminous Ly$\alpha$ emission is common around radio
galaxies, and is thought to be powered by a jet-IGM interaction
\citep[e.g.][]{Chambers90}. Little is known about the source of the
Ly$\alpha$ emission and to what degree the Ly$\alpha$ is scattered by
residual neutral gas. Polarimetry may provide new insights on
the nature of these sources.
\end{itemize}

We have focused on spatially extended Ly$\alpha$ sources, as the polarization averages out to zero for a symmetric unresolved source. However, as was shown by \citet{Lee98}, anisotropic outflows (e.g biconical outflows) may still result in
observable polarization, even if the Ly$\alpha$ source is completely
unresolved. Hence, a measurable polarization may not be restricted to
spatially resolved Ly$\alpha$ sources.

Lastly, we point out that accurate polarization measurements of faint high-redshift sources is difficult. It would therefore be prudent to obtain polarimetry of the brightest sources first. Polarization measurements of one of the brightest known Ly$\alpha$ Blobs (at $z=2.7$, Dey et al. 2005) have already been attempted (Prescott et al. 2008). This approach will likely lead to insights about the processes that regulate the transfer of Ly$\alpha$ in and around high-redshift galaxies. 

\subsection{Comparison to Polarization of Foregrounds}

The dominant foregrounds in the infrared sky (at wavelengths $\lambda <
2\mu$m), are Rayleigh scattered moonlight and starlight
\citep[e.g.][p.33]{Glass99}, and are therefore polarized. 
However, line emitters can easily be extracted
from this scattered foreground for which the spectrum is continuous. Here,
we investigate whether Ly$\alpha$ emitters can be distinguished from {\it
foreground line emitters}, such as [OII], [OIII], H$\alpha$, and H$\beta$
emitters, based on their polarization properties.

In practice it has been possible to effectively discriminate between these
low-redshift interlopers and actual high-redshift Ly$\alpha$ emitters based
on broad-band colors and on the shape of the Ly$\alpha$ line
\citep[e.g.][]{Ka06}. However, the continuum spectrum of some Ly$\alpha$
emitters with a large equivalent width is difficult to detect, and in some
cases the Ly$\alpha$ line is not asymmetric (see e.g the {\it lower left
panel} of Fig~\ref{fig:igm1}).

The possibility that high-redshift Ly$\alpha$ emitters are highly polarized may provide another diagnostic that distinguishes them from low-redshift interlopers. A low level of polarization ($\mathcal{P}\lsim 2\%$) is expected for any extragalactic line emitter due to scattering by dust in the Milky-Way \citep[][]{Schmidt85}. Furthermore, spatially resolved [OIII] line emission at higher levels of polarization ($\mathcal{P}\sim 4\%$) has been observed in a small fraction of Seyfert 2 galaxies \citep{Goodrich92}. This polarization arises when the narrow line region is obscured. In this case, the only [OIII] photons that are observed are scattered towards the observer, either by dust or by electrons. This same mechanism could produce polarized [OII], H$\alpha$ and H$\beta$ line emission. This suggests it is risky to distinguish high-redshift Ly$\alpha$ emitters from low-redshift interlopers on the basis of polarimetry. On the other hand, the observed levels of polarization in low-redshift interlopers are small compared to the values we find in our models, which suggests that high levels of polarization ($\mathcal{P}\sim 10\%$) are indicative of scattered Ly$\alpha$.

Raman scattering of Ly$\beta$ wing photons may produce highly polarized H$\alpha$ emission in a radiative cascade of the form $3p\overset{H\alpha}{\rightarrow}2s\overset{2\gamma}{\rightarrow} 1s$ \citep[e.g.][]{LeeYun,Yoo02}. This mechanism requires the presence of large HI column densities $N_{\rm HI}\gsim 10^{20}$ cm$^{-2}$, and has been observed to operate around symbiotic stars \citep{Ikeda04}, and potentially in close proximity to quasars \citep{LeeYun}. In distant galaxies, however, these emission regions would be completely unresolved, and the polarization averages out to zero. On the other hand, columns in excess of $N_{\rm HI}\gsim 10^{20}$ cm$^{-2}$ can easily exist in the neighborhood of galaxies, and it is not possible to rule out altogether the possibility of highly polarized H$\alpha$ emission as a contaminant of polarized Ly$\alpha$ emitters. Nevertheless, polarized H$\alpha$ lines produced by this mechanism are typically broad (FWHM $\sim 20-40$ \AA) and symmetric, with wings extending into both the longer and shorter wavelengths \citep[see e.g][ their Fig~4-8]{Yoo02}; these features clearly distinguish them from Ly$\alpha$ line emitters.

Ground-based searches for high-redshift Ly$\alpha$ emitters are complicated by the presence of bright OH emission lines, which are produced by the excited radical OH$^{*}$ in the ionosphere, following the reaction H+O$_{3}\rightarrow$ OH$^*$+O$_2$ \citep[e.g.][]{Glass99}. These lines make up the Meinel bands \citep{Meinel50,Meinel50b} and are typically much brighter than Ly$\alpha$ sources. Furthermore, these lines exhibit large temporal and spatial variations in intensity due to the passage of density waves through the ionosphere \citep{Ramsay92}. However, the wavelengths of these lines are well known, and future instruments (such as e.g. the Dark Age Z Ly$\alpha$ Explorer, Horton et al, 2004) intend to search for redshifted ($z=6.5-12$) Ly$\alpha$ lines between these lines using high-resolution ($R=1000$) spectrographs. 

Polarimetry may provide an interesting alternative method of suppressing OH-emission lines. Although the OH-lines have been studied in detail \citep[e.g][]{Ramsay92,Maiharaobs}, we are unaware of any polarization measurements. If OH-lines are unpolarized, then polarimetry may be used to remove them efficiently \footnote{The Meinel bands correspond to rotational and vibrational states of the OH-molecule. It is conceivable that the earth's magnetic field introduces a preferred axis of rotation and/or vibration, which would induce some level of polarization to the lines. However, any large-scale polarization of the foreground OH lines due to the Earth's magnetic field would not have structure on the scales of tens of arcseconds and would therefore be distinguishable from the arc-like features of scattered Ly$\alpha$ radiation around high-redshift galaxies. After removing any uniform polarization component, the remaining foreground should be unpolarized on the small angular scales of interest here. Identifying polarized Ly$\alpha$ lines could in principle be optimized by using special purpose filters that search for the circular polarization pattern shown in Figure~\ref{fig:scheme2} around the sources of interest.}. For example, consider an unpolarized OH sky line that is detected at the $N$--$\sigma$ level after an integration time $t_{\rm int}$. Suppose we repeat the same observation, but we create a first image by passing the incoming radiation through a linear polarizer for $0.5t_{\rm int}$. Next, we create a second image by observing the remaining $0.5t_{\rm int}$ through a linear polarizer that is rotated by 90 degrees compared to the first observation. If one subtracts the two images, then any unpolarized emission is removed down the level of $\sim \sigma$. 

\section{Conclusions}
\label{sec:conc}

The high-redshift Universe is known to contain luminous Ly$\alpha$ emitting
sources such as galaxies and quasars. The Ly$\alpha$ photons that are
emitted by these sources are typically scattered both in the interstellar
medium and in the intergalactic medium. In this paper we have calculated
polarization properties of this scattered Ly$\alpha$ radiation.

We used a Monte-Carlo Ly$\alpha$ radiative transfer code, and endowed each
Ly$\alpha$ photon with a polarization vector which is perpendicular to the
photon's wave vector. In this formulation the Stokes parameters resulted
from binning together multiple independent photons. We differentiated
between resonant and wing scattering, as these are described by different
scattering matrices (see Appendix~\ref{app:super}). Wing scattering is
described by classical Rayleigh scattering and in this case we advanced
the photon's polarization vector in a scattering event following the
well-tested approach of Rybicki \& Loeb (1999). However, resonant
scattering is described by a superposition of Rayleigh scattering and
isotropic scattering with corresponding weights of $1/3$ and $2/3$
(\S~\ref{sec:RT2}). In this case, minor modifications to the approach of
\citet{RL99} are required when advancing the photon's polarization vector
(\S~\ref{sec:RT3}).

We have applied our code to three classes of models which represent the
diverse sets of environments around high-redshift galaxies:

\begin{itemize}

\item First, we computed the polarization of the back-scattered Ly$\alpha$
radiation observed from galaxies surrounded by a superwind-driven outflow.
We have found that the fractional polarization may reach values as high as
$\mathcal{P}_{\rm max}\sim 40\%$, where the maximum polarization depends on
parameters such as the speed of the outflow, $v_{\rm exp}$, and the column
density of neutral hydrogen atoms, $N_{\rm HI}$ (\S~\ref{sec:wind}). In
this case we have found the polarization to increase towards longer
wavelengths.

\item Second, we considered resonant scattering in the intergalactic medium
(IGM) {\it after} reionization. Residual intergalactic hydrogen can scatter up to $\gsim 90\%$ of the Ly$\alpha$ emitted by galaxies out of the line of
sight.  (This high fraction does not apply to galaxies with
superwind-driven outflows where the Ly$\alpha$ photons are scattered out of
resonance before they encounter the IGM.) We have found the polarization of
the scattered Ly$\alpha$ halo around galaxies to be polarized at lower
levels ($\mathcal{P}_{\rm max}\lsim 7\%$) than the halos around galaxies
with outflows. This follows from the fact that Ly$\alpha$ photons scatter
multiple times when they enter resonance, which isotropizes the local Ly$\alpha$ radiation field and reduces its net polarization. The polarization decreases with the mean number of scattering events the photons experience (i.e. the polarization decreases with an increasing effective optical depth in the Ly$\alpha$ line). For this reason, we expect the polarization of resonantly scattered Ly$\alpha$ photons in the IGM to be largest around low redshift sources or around sources where the IGM is more highly ionized than average (e.g. in the vicinity of a bright quasar, see \S~\ref{sec:igm}). 

The overall lower levels of polarization of resonantly scattered Ly$\alpha$ radiation in a reionized IGM are significantly lower than that of pre-reionization Ly$\alpha$ halos around galaxies embedded in a fully neutral IGM \citep{Loeb99}. This suggests that the polarization properties of Ly$\alpha$ halos may help constrain the neutral fraction of the IGM during the epoch of reionization. 

\item Third, we considered neutral collapsing protogalaxies that are
emitting Ly$\alpha$ cooling radiation. Despite the fact that these clouds
are extremely optically thick to Ly$\alpha$ photons, and photons typically
scatter $\sim \tau$ times, we found the polarization to linearly increase
toward the edge of the cloud, reaching a maximum amplitude of
$\mathcal{P}_{\rm max}\sim 35\%$. (\S~\ref{sec:cool}). The resulting
polarization increases towards shorter wavelengths in contrast to the trend
found in outflow models.
\end{itemize}
Our results indicate that resolved high-redshift Ly$\alpha$ emission may be
highly polarized under a variety of likely circumstances. High-redshift
Ly$\alpha$ emitters are usually distinguished from low-redshift line
emitters on the basis of their broad-band colors and their asymmetric
spectral line shape. The work presented here implies that polarization may
provide an additional diagnostic. Moreover, polarimetry has the potential to better remove the glow of infrared lines in the Earth's atmosphere, which would improve the sensitivity of ground-based observations to high-redshift Ly$\alpha$ emitting galaxies outside the currently available redshift windows.

The polarization properties of Ly$\alpha$ radiation encode information
about the distribution and kinematics of neutral gas in and around
galaxies. Polarimetry therefore complements the constraints that are
derivable from spectroscopy. Hence, from a theorist's perspective it is 
well worth to include polarization in Monte-Carlo calculations of
Ly$\alpha$ radiative transfer.

The redshifted Ly$\alpha$ line has provided us with an important
window into the high-redshift Universe. This work suggests that in order to
fully exploit the observations, one should focus on both Stokes
parameters $I$ and $Q$, rather than just $I$. Polarization measurements to better than $1\%$ accuracy can be carried out by existing facilities such as the FOcal Reducer/low dispersion Spectrograph (FORS1)\footnote{http://www.eso.org/instruments/fors/} on the VLT, the LRIS imaging spectropolarimeter at the W.M. Keck Observatory \citep{Goodrich03}, and the CIAO polarimeter on Subaru \citep[][]{Tamura}. The results of this paper imply that Ly$\alpha$ emitting sources would provide excellent future targets for these instruments. 

{\bf Acknowledgments} We thank George Rybicki and Adam Lidz for useful discussions. We thank an anonymous referee for a prompt helpful report that improved the content of this paper. This research was supported by Harvard University funds.

\appendix
\section{Polarization of Resonantly Scattered Ly$\alpha$}
\label{app:super}

\subsection{Ly$\alpha$ Scattering Matrix}

The scattering matrix for a scattering process that is a superposition of
Rayleigh and isotropic scattering of weight $E_1$ and $E_2$ may be written
as \citep[][Eq 250-258]{Chandra60}
\begin{equation}
R= \frac{3}{2}E_1\left( \begin{array}{cc}
\cos^2 \theta & 0 \\
0 & 1 \end{array} \right)+
\frac{1}{2}E_2\left( \begin{array}{cc}
1  & 1 \\
1 & 1 \end{array}\right)
\label{eq:matrix}
\end{equation} where
\begin{equation}
\left( \begin{array}{c}
I_l' \\
I_r' \end{array} \right)=R\left( \begin{array}{c}
I_l \\
I_r \end{array} \right).
\end{equation} 
Here, $\cos \theta={\bf n}\cdot {\bf n}'$ (\S~\ref{sec:RT3}), and $I_l'$ and $I_r'$ are the components of the scattered intensity
parallel and perpendicular to the plane of scattering, respectively. If the
incoming light is assumed to be unpolarized, then $I_l=I_r=\frac{1}{2}$,
and the phase function and degree of polarization are given by
\citep[][Eqs. 250--258]{Chandra60}
\begin{equation}
p(\theta)=E_2+\frac{3}{4}E_1+\frac{3}{4}E_1\cos^2 \theta, \hspace{2mm}
\Pi(\theta)=\frac{\sin^2\theta}{1+\frac{4E_2}{3E_1}+\cos^2\theta}
\end{equation} 
Comparison with Eq.~(\ref{eq:phasecore}) reveals that Ly$\alpha$ resonant
scattering of unpolarized light corresponds to the case $E_1=\frac{1}{3}$
and $E_2=\frac{2}{3}$.

As was mentioned in \S~\ref{sec:RT2}, the non-zero degree of polarization
arises fully\footnote{A simple way to see why scattering through the
$2P_{1/2}$ level does not result in any polarization is that the angular
part of the wavefunction describing the electron in the $2P_{1/2}$-state is
a constant \citep[e.g.][p. 134]{White34}. Hence an atom that is excited
into the $2P_{1/2}$-state has 'lost' any memory of the direction and/or
polarization of the photon that excited the atom. The photon that is
emitted when the atom decays back to the ground state has no preferred
direction (i.e. it is emitted with an equal probability in all
directions).}  from scattering between the levels $1S_{1/2}\rightarrow
2P_{3/2} \rightarrow 1S_{1/2}$. This transition is denoted by H (while K
denotes the $1S_{1/2}\rightarrow 2P_{1/2} \rightarrow 1S_{1/2}$
transition). \citet{Hamilton47} showed that for the H and K transitions
$(E_1,E_2)_{\rm H}=(\frac{1}{2},\frac{1}{2})$ and $(E_1,E_2)_{\rm
K}=(0,1)$. Hence, scattering by 90$^{\circ}$ via the H-transition results in
$\Pi(90^{\circ})=\frac{3}{7}$, as was already mentioned in
\S~\ref{sec:RT2}. Because scattering is twice as likely to occur via the H
transition, the scattering averaged value becomes
$(E_1,E_2)=\frac{1}{3}(E_1,E_2)_{\rm K}+\frac{2}{3}(E_1,E_2)_{\rm
H}=(\frac{1}{3},\frac{2}{3})$.

Lastly, we point out that Rayleigh and isotropic scattering by $90^{\circ}$
results in $100$ and $0\%$ polarization, respectively. Since resonant
scattering may be viewed as a superposition of Rayleigh scattering and
isotropic scattering with corresponding weights of $1/3$ and $2/3$, one
might expect $90^{\circ}$ resonant scattering to result in a fractional polarization of $\frac{1}{3}$, while the actual number is
$\Pi(90^{\circ})=\frac{3}{11}<\frac{1}{3}$. The main reason for this difference
is that isotropic and Rayleigh scattering have different phase functions,
and that {\it if} scattering by $90^{\circ}$ occurs, then the probability
that this was due to isotropic scattering is larger than $\frac{2}{3}$. Quantitatively, the angular dependence of the fractional contribution of Rayleigh scattering to the phase-function is given by $p_R(\theta)=\frac{1}{3}(\frac{3}{4}+\frac{3}{4}\cos^2\theta)/(\frac{11}{12}+\frac{1}{4}\cos^2\theta)=({1+\cos^2\theta})/(\frac{11}{3}+\cos^2\theta)$, which was used in \S~\ref{sec:RT3}.

\subsection{Quantum Interference as in Stenflo (1980)}

\begin{figure}
\vbox{\centerline{\epsfig{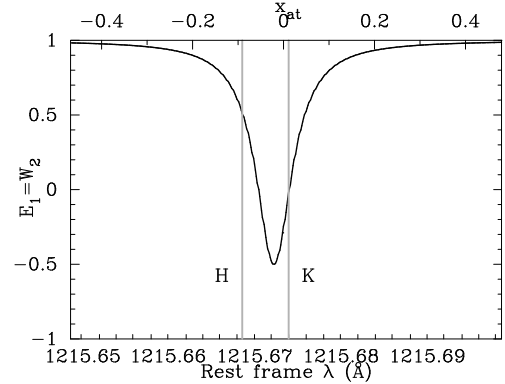}}}
\caption[]{The impact of quantum interference on the polarization in
resonant Ly$\alpha$ scattering. The frequency dependence of the parameter
$E_1$ (see Eq.~\ref{eq:matrix}), as calculated by \citet{Stenflo80}, is
shown. One finds $E_1=\frac{1}{2}$ and $E_1=0$ at the H and K resonance
frequencies as expected. At sufficiently large separations from both
resonances, $E_1 \rightarrow 1$, which corresponds to pure Rayleigh
scattering. Interestingly, $E_1<0$ for a range of frequencies between the H
and K resonance frequency. In practice, however, scattering at these
frequencies does not occur often enough to leave an observable imprint (see
text, and Fig~\ref{fig:app2}). The upper horizontal axis shows frequency in
normalized units $x$ (we arbitrarily set $x=0$ at $\nu=\nu_{\rm K}$).}
\label{fig:app}
\end{figure}

In \S~\ref{sec:RT2} the phase function and the degree of polarization
caused by resonant and wing scattering were given. Here we discuss the
transition between these two regimes. \citet{Stenflo80} has calculated the
value of $E_1$ as a function of frequency (but note that $E_1=W_2$ in the
notation of Stenflo 1980, Eq.~3.25). We have plotted $E_1$ as a function of
wavelength (bottom label) and frequency (top label) in
Figure~\ref{fig:app}.

Figure~\ref{fig:app} shows that $E_1=\frac{1}{2}$ at the H resonance
frequency and $E_1=0$ at the K resonance frequency, which correspond
exactly to the values quoted above. The Figure also shows that at
sufficiently large separations from both resonances, $E_1 \rightarrow
1$. In dimensionless frequency units, this asymptotic value is practically
reached when $|x_{\rm at}|\gsim 0.2$. This was the motivation for using the
threshold $x_{\rm crit}=0.2$ to separate core from wing scattering in
\S~\ref{sec:RT3}.

\begin{figure}
\vbox{\centerline{\epsfig{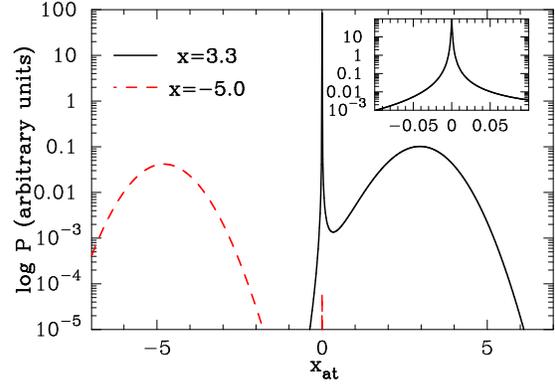}}}
\caption[]{The probability that a photon of frequency $x$ is scattered by
an atom such that it appears at a frequency $x_{\rm at}$ in the frame of
the atom. The {\it solid} and {\it dashed lines} correspond to $x=3.3$ and
$x=-5.0$ respectively, $P(x_{\rm at},x)$ (Eq.~\ref{eq:u3}). For $x=3.3$,
photons are either scattered by atoms to which they appear {\it exactly} at
resonance (see inset), or to which they appear $\sim 3$ Doppler widths
away. For $x=-5$, resonant scattering is less important by orders of
magnitude. In combination with Fig~\ref{fig:app}, this figure shows that if
a photon is resonantly scattered then $E_1$ is either $0$ or
$\frac{1}{2}$.}
\label{fig:app2}
\end{figure}

Interestingly, $E_1<0$ for a range of frequencies between the H and K
resonance frequencies. However, scattering at these frequencies does not
occur often enough to leave an observable imprint. The reason for this is
simple: the natural width of the line for both the H and K transitions is
much smaller than their separation, i.e $\gamma_{\rm H,K}\sim 10^8$ Hz $\ll
\nu_H-\nu_K=1.1\times 10^{10}$ Hz. Since the absorption cross--section in
the atom's rest-frame scales as, $\sigma(\nu)\propto [(\nu-\nu_{\rm
H,K})^2+\gamma_{\rm H,K}^2/4]^{-1}$, a Ly$\alpha$ photon is much more
likely absorbed by an atom for which the photon appears exactly at
resonance, than by an atom for which the photon has an energy corresponding
to a negative $W_2$. Quantitatively, the Maxwellian probability $P$ that a
photon of frequency $x$ is scattered at frequency $x_{\rm at}$ in the
atom's rest-frame is given by
\begin{equation} P(x_{\rm at},x)dx_{\rm at}=\frac{a}{\pi
H(a,x)}\frac{e^{-(x_{\rm at}-x)^2}}{x_{\rm at}^2+a^2} dx_{\rm at},
\label{eq:u3}
\end{equation} 
where $H(a,x)$ is the Voigt function ($H(a,x)\equiv \tau_x/\tau_0$, see Eq~\ref{eq:phi}). This probability is shown in Figure~\ref{fig:app2} for\footnote{Note that the transition from core to wing scattering occurs at $x \sim 3.3$, see
\S~\ref{sec:RT1}.} $x=3.3$ ({\it solid line}) and $x=-5.0$ ({\it dashed
line}).  Figure~\ref{fig:app2} shows that for $x=3.3$, photons are
scattered either when they are exactly at resonance or when they appear
$\sim 3$ Doppler widths away from resonance. The inset of
Figure~\ref{fig:app2} zooms on the region near $x_{\rm at}=0.0$. Clearly,
if a photon is resonantly scattered, then it is more likely to be scattered
when it is {\it exactly} at resonance then when it is, say, $0.05$ Doppler
widths away from resonance. For frequencies $|x|>3.3$ there are not enough
atoms moving at velocities such that the Ly$\alpha$ photon appears at
resonance in the frame of the atom. Instead, the photon is scattered while
it is in the wing of the absorption profile. This is illustrated by the
{\it dashed line} which shows the case $x=-5.0$, for which resonant scattering is less likely by orders of magnitude.

From the above we conclude that ({\it i}) when a photon is scattered in the
wing of the line profile, then $E_1=1$ and ({\it ii}) when a photon is
scattered resonantly, then it is scattered almost exactly at resonance,
and therefore $E_1=\frac{1}{3}$ (the scattering-averaged value).

\subsection{A Note on the Polarization Dependence of the Scattering Phase-Function}
\label{app:phase}

In this section we show that the procedure for propagating the wave and polarizations vector in a single Rayleigh scattering event that was outlined in \S~\ref{sec:RT3}, naturally accounts for the polarization dependence of $p(\theta)$ and $\Pi(\theta)$.
\begin{figure}
\vbox{\centerline{\epsfig{file=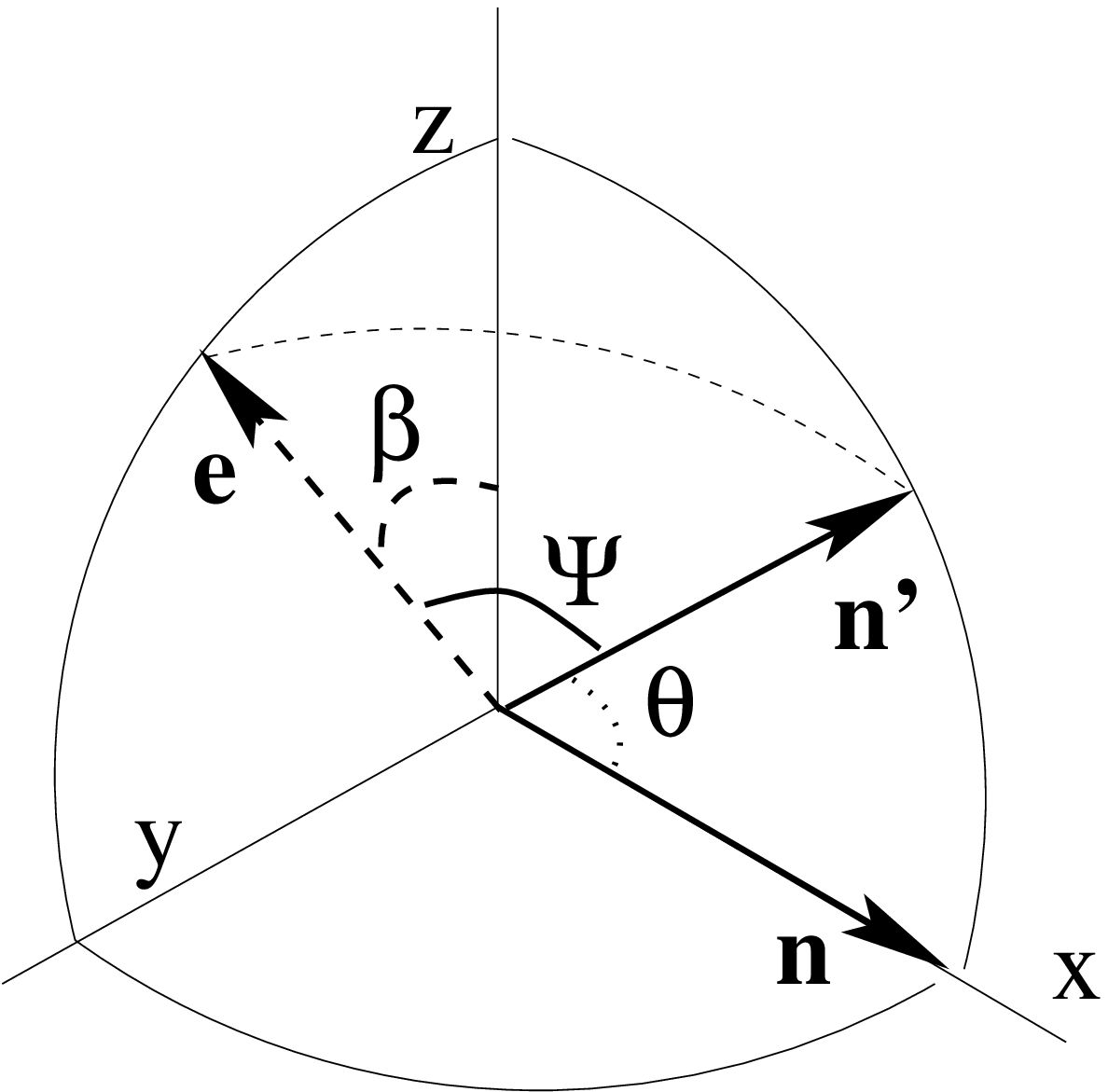,angle=0,width=4.0truecm}}}
\caption[]{The coordinate system that is used to investigate the dependence of the scattering phase function on the polarization of the incoming radiation.}
\label{fig:app3}
\end{figure}

We introduce a rectangular coordinate system in which a Ly$\alpha$ photon propagates along the $x-$axis prior to scattering, and in the $x-z$ plane after scattering (see Fig~\ref{fig:app3}). In this coordinate system, ${\bf n}=(1,0,0)$, ${\bf n'}=(\cos \theta,0,\sin \theta)$, and ${\bf e}=(0,\sin \beta, \cos \beta)$, where the notation of \S~\ref{sec:RT3} was used. Here, $\beta$ denotes the angle between the polarization vector ${\bf e}$ and the $x-z$ plane, and $\Psi$ denotes the angle between ${\bf e}$ and ${\bf n'}$. The components of the intensity parallel and perpendicular to the plane of scattering -prior to scattering- are given by $I_{\rm l}=I \cos^2 \beta $ and $I_{\rm r}=I \sin^2 \beta $, respectively. From the scattering matrix it follows that the phase function for this polarized radiation is given by
\begin{equation}
p(\theta)d\Omega=\Big{(}\frac{3}{2}\cos^2\theta \cos^2\beta+\frac{3}{2}\sin^2\beta\Big{)}d\Omega.
\end{equation} Using that $\cos \Psi= {\bf n}'\cdot {\bf e}=\sin \theta \cos \beta$ we obtain

\begin{equation}
p(\Psi)d\Omega=\frac{3}{2}\sin^2\Psi d\Omega.
\end{equation} Hence, the probability that a photon is emitted into solid angle $d\Omega(\theta,\phi)$ scales simply as $\sin^2\Psi$ (where $\Psi=\Psi(\theta,\phi)$). This phase-function was used in this paper for Rayleigh scattering ($\sin^2\Psi=1-({\bf e}\cdot {\bf n'})^2$). Hence, this method naturally incorporates the polarization dependence of scattering phase function. Furthermore, the intensity parallel and perpendicular to the plane of scattering after scattering are given by $I'_{\rm l}=\frac{3}{2}I \cos^2 \beta \cos^2 \theta$ and $I'_{\rm r}=\frac{3}{2}I \sin^2 \beta $, respectively. The polarization vector of this scattered radiation is therefore indeed given by ${\bf e'}$, which was obtained by projecting ${\bf e}$ onto the plane perpendicular to ${\bf n}'$ (see \S~\ref{sec:RT3}). 

\label{lastpage}
\end{document}